\documentclass[12pt]{article}

\usepackage[a4paper,left=30mm,right=20mm,top=20mm,bottom=20mm]{geometry}
\usepackage{graphics}
\usepackage{amsmath}
\usepackage{amssymb}
\usepackage{epsfig}
\usepackage{cite}
\usepackage{xcolor}
\usepackage[english]{babel}

\def\MGvATNLO{{\sc MadGraph5\_aMC@NLO}}
\newcommand{\MGTMDZj}{{\sc MG5\-aMC+CA3}(Z+1)NLO}
\newcommand{\cas}{{\sc Cascade3}}

\title{Discriminating the heavy jet production mechanisms in associated $Z$ + heavy flavor events at the LHC}

\author{S.P.~Baranov$^{1}$, 
A.~Bermudez Martinez$^{2}$, 
H.~Jung$^{2}$, \\ A.V.~Lipatov$^{3,4}$, M.A.~Malyshev$^{3}$, S.~Taheri~Monfared$^{2}$ }

\begin{document}


\maketitle

\vspace*{-7.5cm}
\begin{flushright}
DESY-21-184
\end{flushright}
\vspace*{+4.5cm}

\vspace{0.5cm}

\begin{center}

{\it $^{1}$Lebedev Physics Institute, Moscow, Russia}\\
{\it $^{2}$Deutsches Elektronen-Synchrotron DESY, Germany}\\
{\it $^{3}$Skobeltsyn Institute of Nuclear Physics, Lomonosov Moscow State University, Moscow, Russia}\\
{\it $^{4}$Joint Institute for Nuclear Research, 141980, Dubna, Moscow region, Russia}\\

\end{center}

\vspace{0.5cm}

\begin{center}

{\bf Abstract }

\end{center}

\indent

We reconsider the associated $Z$ boson and charm or beauty jet production at the LHC with paying special attention
to the formation dynamics of heavy jets. 
Two different approaches are studied: first one, where heavy quarks are produced 
in the hard scattering subprocesses, implemented in the Monte-Carlo generator \textsc{pegasus},
and another scenario, where the hard scattering is calculated at NLO with \MGvATNLO\ and TMD parton shower is included (implemented in the Monte-Carlo generator \cas ).
We compare the predictions obtained in both  schemes 
with latest experimental 
data for associated $Z + b$ production cross sections and the relative production 
rate $\sigma(Z + c)/\sigma(Z + b)$ collected by the ATLAS and CMS Collaborations at $\sqrt s = 13$~TeV.
We introduce two kinematic observables (denoted as 
$z_b$ and $p_T^{\rm rel}$) which can be used to discriminate the heavy jet production 
mechanisms. Using these variables we trace the shape of the simulated $b$-jet events
and recommend that these observables be taken into consideration in the forthcoming 
experimental analyses.

\vspace{1.0cm}


\newpage
\section{Motivation} \indent

With the LHC in operation, one can access a number of `rare' processes which could have 
never been systematically studied at previous accelerators. In this article we revisit 
the associated production of $Z$ bosons and $b$-jets. This process involves both weak and 
strong interactions and therefore serves as a complex test of the Standard Model, perturbative 
QCD and our knowledge of parton densities. On the experimental side, we have at our disposal 
the data collected by ATLAS and CMS collaborations\cite{CMS7,CMS8,CMS13,ATLAS7,ATLAS13}.

Earlier, we have demonstrated\cite{combined,combined2} a quite reasonable agreement between 
the theoretical and experimental results with respect to many observables, such as the
differential cross sections and particle correlations.
Here we wish to go one step deeper in our understanding and draw attention to two new
observables which can be used as clean probes of $b$-jet formation dynamics.
We aim at a difference between `prompt' and `non-prompt' production cases.
The former class refers to the situation when the $b$-quark
is produced in the hard scattering subprocess; it further radiates lighter partons
and after all evolves into a jet containing $b$-hadrons. The latter class refers to jets 
initiated by a light parton (gluon or quark), and then $b$-quarks (or rather $b$-hadrons) 
appear from a parton evolution cascade.
In principle, one can distinguish between these two kinds of $b$-jets 
using an appropriate machine learning technique for gluon-quark jet classification. 
However, this is not straightforward and should be done carefully\cite{Pumplin,quark-gluon}. 

We discuss kinematic criteria that can be helpful to discriminate
these cases. Namely, we find that in the jets originating from $b$-quarks (i.e., `prompt'),
$b$-hadrons carry larger momentum fraction $z_b$ than in other (`non-prompt') jets. 
Second, in the prompt jets, $b$-hadrons move closer to the jet axis. 
The goal of the study is to give quantitative estimates and to see to what extent our 
expectations are supported by real data. 

\section{Theoretical framework} \indent

There are two commonly used approaches in perturbative QCD (pQCD) calculations 
for cross sections of processes containing heavy quarks. 
One of these approaches is the so-called four-flavour number scheme (4FNS), where 
only gluon distributions and first two quark generations are involved 
in the QCD evolution equations for parton (quark and gluon) densities in a proton, so that
$b$-quarks appear in a massive final state as a result of gluon splitting $g\to b\bar b$.
The second approach is the five-flavour number scheme (5FNS), which allows 
a $b$-quark density in the initial state where the $b$-quark is typically treated massless
above the flavour threshold.
Up to all orders, the 4FNS and 5FNS schemes should give exactly the same results,
while at a given order difference can occur%
\footnote{The discussion on the advantages and disadvantages of the different flavour 
number schemes can be found, for example, in review\cite{4FNSvs5FNS}. The consistency of both approaches 
within the context of Parton Branching (PB) approach\cite{PB1,PB2} has been recently discussed\cite{4FNSvs5FNS-PB}.}.

\subsection{Calculations with Monte-Carlo generator PEGASUS} \indent

To calculate the $Z+b$-jet total and differential cross sections at the LHC we employ 
two different schemes based on the transverse momentum dependent (TMD) 
quark and gluon distributions in a proton\footnote{For detailed description and 
discussion of the different approaches involving TMD parton densities see, for example, review\cite{kt-rev}.}. 
The first method was proposed in\cite{combined}
and relies mainly on the ${\cal O}(\alpha \alpha_s^2)$ off-shell gluon-gluon fusion subprocess:
\begin{gather}
  g^* + g^* \to Z + b + \bar b,
\label{ggZ}
\end{gather}
\noindent
which gives the leading contribution to the production cross section
in the small $x$ region, where the gluon density dominates over the quark densities.
An essential point here is using the CCFM evolution equation to describe the QCD 
evolution of the TMD gluon density in a proton.
The gauge-invariant off-shell amplitude for the gluon-gluon fusion subprocess has been 
calculated in Ref.~\cite{ggZbb1,ggZbb2}, where all the relevant technical details are explained.

In addition, we take into account two subleading subprocesses involving 
quarks in the initial state. These are the flavor excitation processes
\begin{gather}
  q + Q \to Z + Q + q,
\label{qqtZ}
\end{gather}
\noindent
and the quark-antiquark annihilation processes
\begin{gather}
  q + \bar q \to Z + Q +\bar Q,
\label{qqsZ}
\end{gather}
\noindent
which could play a role essentially at large transverse momenta 
(or, respectively, at large $x$ which is needed to produce large $p_T$ events)
where the quarks are less suppressed or can even dominate over the gluon density.
The contributions from the quark-induced subprocesses~(\ref{qqtZ}) and (\ref{qqsZ}) are
calculated within a conventional DGLAP-based (collinear) factorization scheme, which provides 
better theoretical grounds in the region of large $x$. The evaluation of the production 
amplitudes is straightforward and needs no explanations. 

Our scheme\cite{combined,combined2} represents a combination of two techniques
with each of them being used at the kinematic conditions where it is best suitable.
This scheme is implemented in the Monte-Carlo event generator \textsc{pegasus}\cite{pegasus}, 
which has been used for numerical calculations.
Taking all the three subprocesses~(\ref{ggZ}), (\ref{qqtZ}) and (\ref{qqsZ})
into account we extend the predictions to the whole kinematic range.
Note that at least one heavy quark $Q$ is always present in the final state already 
at the amplitude level.

The parton-level calculation returned by \textsc{pegasus} has further been improved by 
including the effects of the initial and final state parton showering. 
For the collinear part of the calculation, that has been done using the 
conventional \textsc{pythia8}\cite{pythia} 
algorithm
\footnote{In fact, we took the TMD parton shower tool implemented into the Monte Carlo event generator \protect\cas\ ~\cite{cascade} and applied it to the off-shell gluon-gluon fusion subprocess~(\ref{ggZ}).}. 
The off-shell part of the calculations includes this kind of correction in the form
of TMD gluon densities.
The subsequent decay $Z \to l^+l^-$ (including the $Z/\gamma^*$ interference effects) 
is incorporated already at the production step at the amplitude level in order to fully 
reproduce the experimental setup.

For the TMD gluon density in a proton, we used a numerical solution of the CCFM
equation\cite{CCFM1,CCFM2,CCFM3,CCFM4}.
We find it to be a suitable option since it smoothly interpolates between the 
small-$x$ Balitsky-Fadin-Kuraev-Lipatov (BFKL)\cite{BFKL1,BFKL2,BFKL3} gluon dynamics
and large-$x$ DGLAP one. We adopted the latest JH-2013 parametrization, 
namely, we choose the JH-2013 set 2\cite{JH2013}. 
The corresponding TMD gluon density was fitted to high-precision DIS data on the proton 
structure functions $F_2(x, Q^2)$ and $F_{2}^c(x,Q^2)$. The fit was based on TMD
matrix elements and involves the two-loop strong coupling
constant, the kinematic consistency constraint\cite{const1,const2} and
non-singular terms in the CCFM gluon splitting function\cite{CCFM-split}.
For the conventional quark and gluon densities, we used
the MMHT'2014 (NLO) set\cite{mmht}.

\subsection{Calculations with Monte-Carlo generator \bf\cas } \indent

This method represents a more rigorous scheme based on the parton branching (PB) approach,
which was introduced\cite{PB1,PB2} to treat the DGLAP evolution\cite{DGLAP1,DGLAP2,DGLAP3,DGLAP4}. The method provides a
solution of these equations and coincides with the standard methods to solve the 
DGLAP equations for inclusive distributions at NLO and NNLO. It allows one
to simultaneously take into account soft-gluon emission at $z\to 1$ 
and the transverse momentum $q_T$ recoils in the parton branchings along the QCD cascade. 
The latter leads to a natural determination of the TMD quark and gluon densities in a proton.
One of the advantages of this approach is that the PB TMDs can be combined with standard (on-shell) production amplitudes, which can be calculated at higher orders with. Here we use matrix elements calculated with next-to-leading (NLO) order with  \MGvATNLO\  \cite{madgraph} using the HERWIG6 subtraction terms, which are suitable for combination with PB-TMDs.

The tool to be used to calculate the observables within the PB approach is the Monte-Carlo 
event generator \cas ~\cite{cascade}. A special procedure is adopted for
the initial partons' transverse momenta: a transverse momentum is
assigned according to the TMD density, and then the parton-parton system is boosted 
to its center-of-mass frame and rotated in such a way that only the longitudinal and 
energy components are nonzero. The energy and longitudinal component of the initial momenta 
are recalculated taking into account the virtual masses\cite{cascade,BSvZ}. 
This method keeps the parton-parton invariant mass exactly conserved, while 
the rapidity of the partonic system is approximately restored, depending on the transverse
momenta.

The PB TMD parton densities were obtained via fitting to precise HERA DIS data. 
Two sets of them, which differ from each other by the choice of the scale in the QCD
coupling, were obtained~\cite{PB-TMD}. In the numerical calculations below 
we use\footnote{All the TMD quark and gluon densities in a proton used here
are available via the TMDlib tool~\cite{tmdlib}.} the PB-NLO-HERAI+II-2018 set 2. 

\section{Numerical results} \indent

Before we turn to the discussion on the 'prompt' and 'non-prompt' $b$-jets we have
to justify our approach by confronting the results of our simulations with the newest 
CMS and ATLAS data. We start by listing the parameters of our calculations. 
Following\cite{PDG}, we set the charm and beauty quark masses to
$m_c = 1.4$~GeV and $m_b = 4.75$~GeV, the mass of $Z$ boson $m_Z = 91.1876$~GeV, 
its total decay width $\Gamma_Z = 2.4952$~GeV, and $\sin^2\theta_W = 0.23122$. The 
electromagnetic coupling is taken as $\alpha(m_Z)=1/128.74$.

In  the calculations performed with \textsc{pegasus} we set 
$\Lambda_{\rm QCD}^{(4)} = 200$~MeV and use two-loop QCD coupling according 
to\cite{JH2013}.
The default renormalization scale was taken as $\mu_R^2 = m_Z^2$, while
the default factorization scale for the off-shell gluon-gluon fusion 
subprocess was $\mu_F^2 = \hat s + {\mathbf Q}_T^2$,
where ${\mathbf Q}_T$ is the net transverse momentum of the initial off-shell gluon 
pair. This choice is dictated mainly by the CCFM evolution algorithm (see\cite{JH2013} 
for more information). For the quark-induced subprocesses~(\ref{qqtZ}) and (\ref{qqsZ})
we keep it equal to the renormalization scale, $\mu_F=\mu_R$. 
To estimate the theoretical uncertainties of our \textsc{pegasus} 
calculations for off-shell gluon-gluon fusion subprocess~(\ref{ggZ})
we use auxiliary '+' and '$-$' TMD gluon densities 
instead of the default TMD gluon distribution functions.
These two sets refer to the varied hard scales in the strong coupling constant $\alpha_s$ in the off-shell amplitude: '$+$' 
stands for $2\mu_R$, while '$-$' refers to $\mu_R/2$ (see\cite{JH2013}). 
For the quark-induced subprocesses~(\ref{qqtZ}) and (\ref{qqsZ}) we just vary the hard scale around its 
default value between halved and doubled magnitude, as usual.

For the PB calculation with \MGTMDZj , we set $m_c = 1.47$~GeV, $m_b = 4.5$~GeV, $\alpha_s (m_Z^2) = 0.118$ and $\mu_R = \mu_F = \frac{1}{2} \sum_i \sqrt{m^2_i +p^2 _{t,i}}$,  where the sum runs over all particles and parton in the matrix element.  The hard process calculations are performed at NLO with \MGvATNLO\  ~\cite{madgraph} with  \textsc{herwig6} subtraction terms and the TMD parton shower is simulated with \cas ~\cite{cascade}. The theoretical uncertainties are obtained by varying the scale of the hard process is varied by a factor 2 up and down, provided by \MGvATNLO .

\subsection{$Z$ + heavy quark jet production} \indent

Now we are in position to present our numerical results.
First we discuss $Z + b$ production cross sections measured  by the 
ATLAS Collaboration at $\sqrt{s} = 13$~TeV\cite{ATLAS13}. 
The experimental setup was as following: the transverse momenta of the leptons from the $Z$ decay
are required to be $p_T^\text{lead}>27$, with pseudorapidities  of $|\eta^l|<2.5$ for 
muons or $|\eta^l|<2.47$ for electrons (excluding $1.37<|\eta^l|<1.52$). 
The leptons are isolated from the jets by $\Delta R<0.4$. The invariant mass of the 
reconstructed $Z$-boson has to be within $76<m^{ll}<106$. The jets are reconstructed 
in the anti-$k_T$ algorithm with radius $R_\text{jet}=0.4$ and should have 
$p_T^\text{jet}>20$~GeV and rapidity $|y|<2.5$. 

Our numerical predictions are shown in Figs.~1 and 2 in comparison with the latest ATLAS 
data\cite{ATLAS13}. The shaded bands represent the theoretical uncertainties estimated as described above. In all figures we show predictions (as described above) obtained from \MGTMDZj\ based on a NLO calculation of $Z+1$ jets with \MGvATNLO\ and TMD parton showers from \cas\ with and without hadronization, together with predictions obtained from \textsc{pegasus} with and without parton shower.

The \MGTMDZj\  calculations describe well the ATLAS measurements of the $Z$ boson and the $b$ jet rapidity and transverse momenta spectra at low and moderate $p_T$
within the theoretical and experimental uncertainties. From the comparison of \MGTMDZj\ with and without fragmentation, we estimate the fragmentation correction of $\sim 10 \% $ in the larger transverse momentum regions, while at small $p_T(Z)$ and small $p_T(bjet)$ the corrections are significantly larger. These corrections are coming from $b$-hadrons which are outside the jet with $R=0.4$.
Due to missing  higher order contributions in the calculation ($Z+1$ at NLO) there is a notable underestimation of the data at large transverse momenta, namely $p_T \gtrsim 200$~GeV.
These missing higher order contributions lead also to deviations at small $\Delta\phi(Z,b)$, $\Delta R(Z,b)$ and  at large $\Delta y(Z,b)$.


The \textsc{pegasus} predictions  describe the data quite well within the estimated uncertainties, failing though at large $p_T$. 
The scenario implemented into \textsc{pegasus},
where the heavy quark is produced in the hard partonic 
scattering at the amplitude level, is able to reproduce well
the measured distributions in $Z$ boson and $b$ jet rapidity and transverse momenta (at low and moderate $p_T$).
It is interesting to observe, that the distribution of $\Delta\phi(Z,b)$ is well described, even at low $\Delta\phi(Z,b)$, which is in contrast with the PB result.
One should however keep in mind that the \textsc{pegasus} calculations do not include fragmentation (unlike the PB ones). Taking into account the fragmentation effect may result in a $\sim 10$\% drop of cross section~\cite{had_eff}.
The inclusion of the corresponding fragmentation correction factor could simultaneously result in better agreement by the two approaches.
In the $\Delta y(Z,b)$ we observe a similar behavior as for the PB predictions.
The final state parton shower effects does not significantly affect the \textsc{pegasus} predictions.
It can be easily understood since the main contributions here
comes from initial state gluon radiation, which is already taken into 
account in the CCFM-evolved gluon density.

Now we turn to associated $Z + c$ production at the LHC and 
discuss the relative production rate $\sigma(Z + c)/\sigma(Z + b)$
as measured by the CMS collaboration at $\sqrt{s}=13$~TeV\cite{CMS13}. 
The experimental cuts are: the leading lepton from the $Z$ decay is required to have a transverse 
momentum $p_T^\text{lead}>26$~GeV, while the subleading lepton must have $p_T^\text{sublead}>10$~GeV; 
with pseudorapidities within $|\eta^l|<2.4$ and the dilepton invariant mass should be $71<m^{ll}<111$. 
The leptons are required to be isolated  from the jets with $\Delta R<0.4$.
The jets are required to have $p_T^\text{jet}>30$~GeV and $|\eta^{\text {jet}}|<2.4$, reconstructed 
with the anti-$k_T$ algorithm with radius $R_\text{jet}=0.4$.

In Fig.~3 we show the results of our calculations 
for the ratio $\sigma(Z + c)/\sigma(Z + b)$ as a function 
of $Z$ boson or jet transverse momenta 
in comparison with the CMS data\cite{CMS13}.
The shaded bands represent the 
estimated uncertainties of our calculations.
A good description of the CMS data can be 
obtained with \MGTMDZj\ 
The predictions from \textsc{pegasus} lie
below the data, though being compatible with the data at $\sim2\sigma$. 
Thus, despite the fact that both considered scenario provide a more or less 
reasonable description of inclusive production data (see Figs.~1 and 2),
they are very distinct for $\sigma(Z+c)/\sigma(Z+b)$ observables 
and some correlation variables.
The same conclusion was achieved earlier\cite{combined2} 
when comparing the \textsc{pegasus} and \textsc{sherpa} (NLO pQCD) predictions, 
where the same heavy jet production scenario, as in the \MGTMDZj\, is employed.
Below we will discuss the kinematic criteria that can be helpful in 
discriminating these two cases. 

\begin{figure}
\begin{center}
\includegraphics[width=7.8cm]{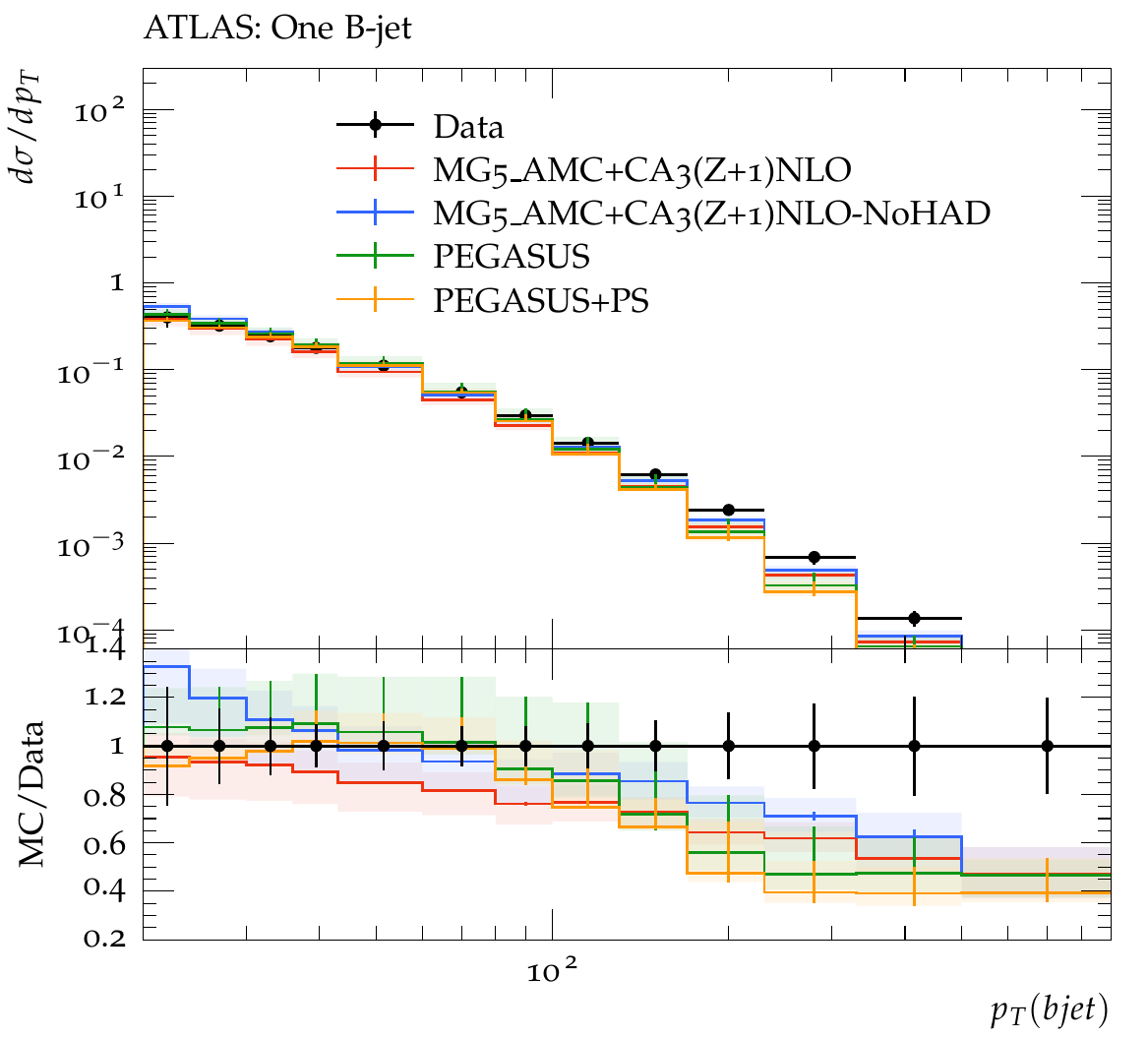}
\includegraphics[width=7.8cm]{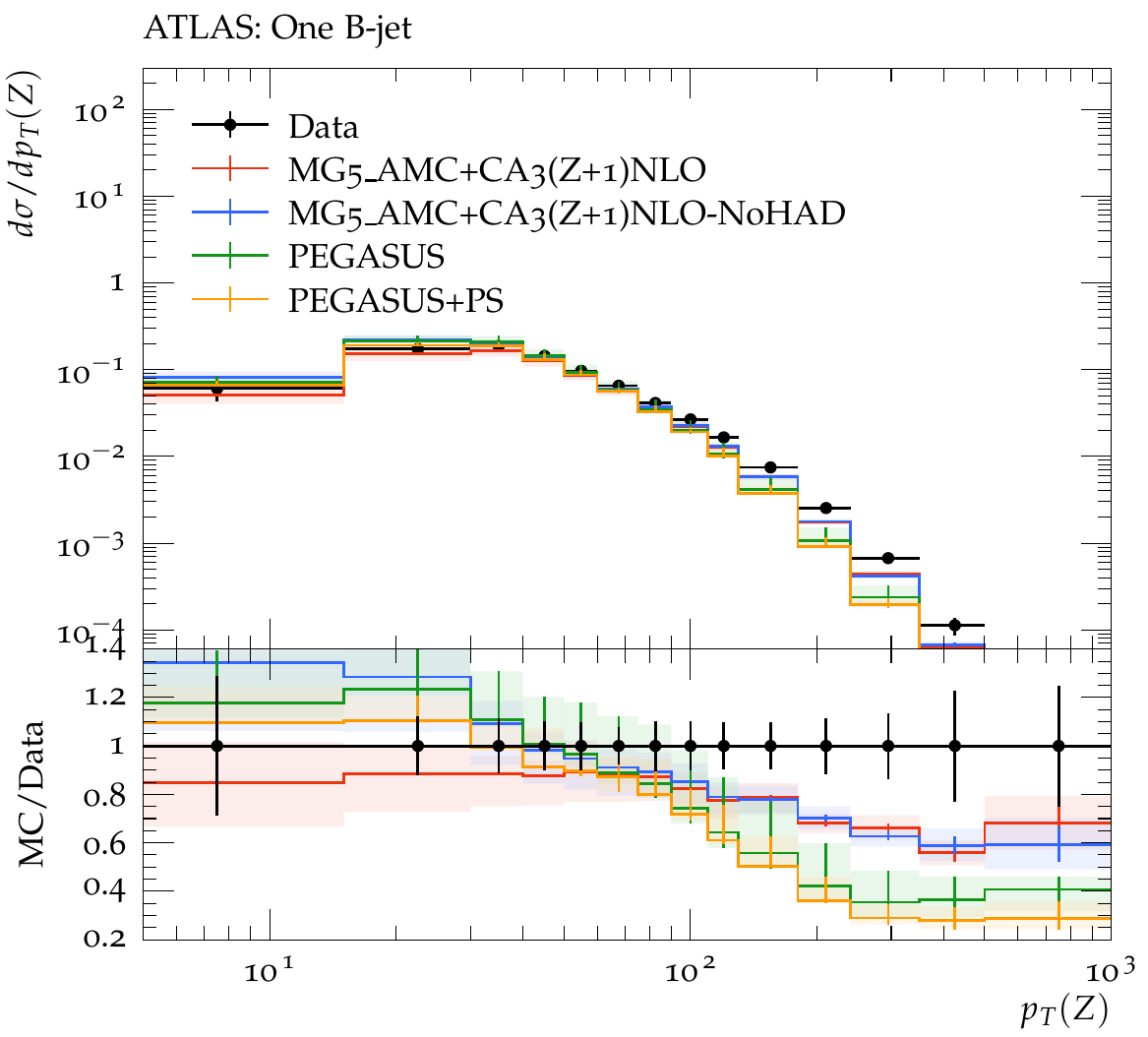}
\includegraphics[width=7.8cm]{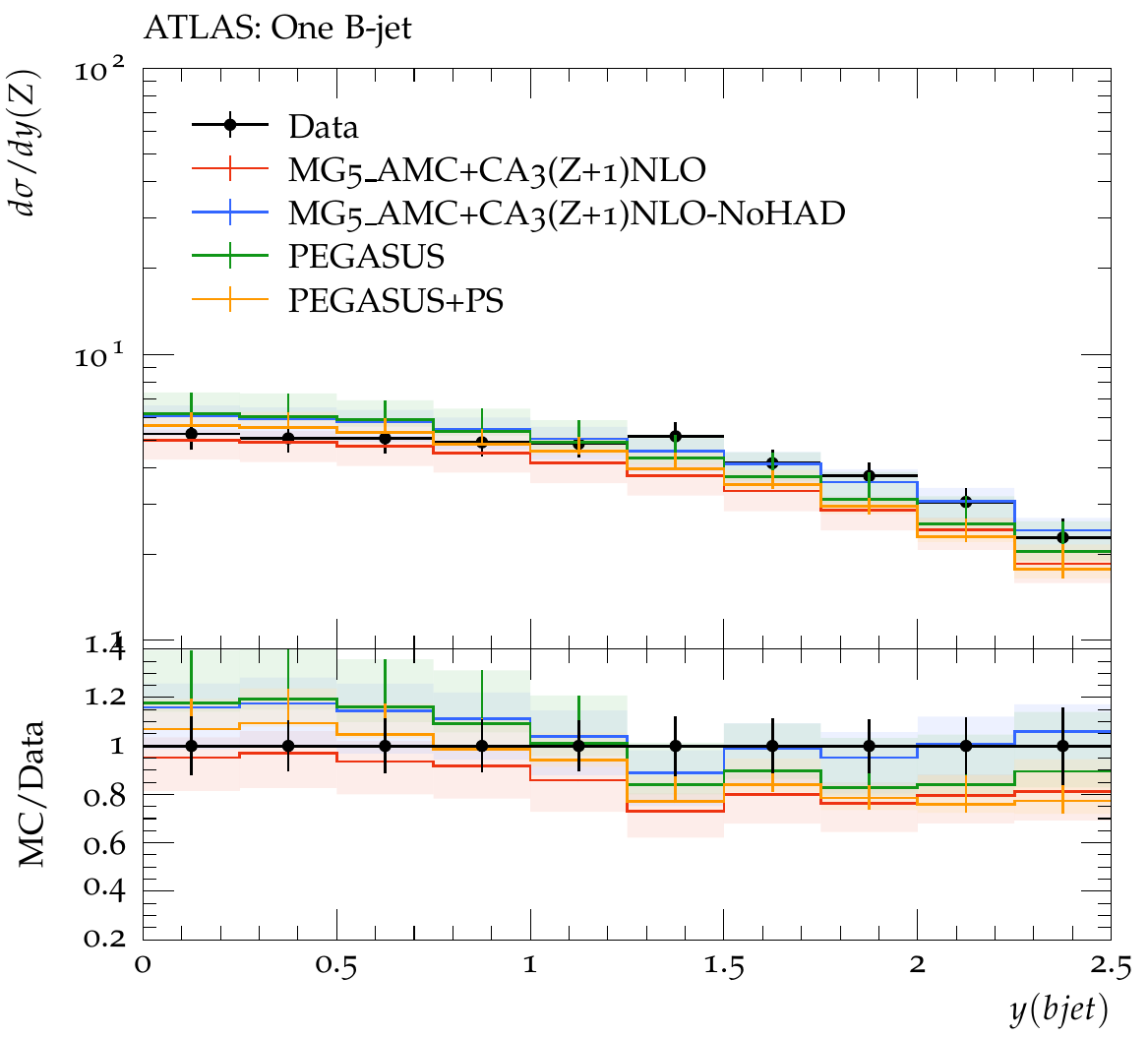}
\includegraphics[width=7.8cm]{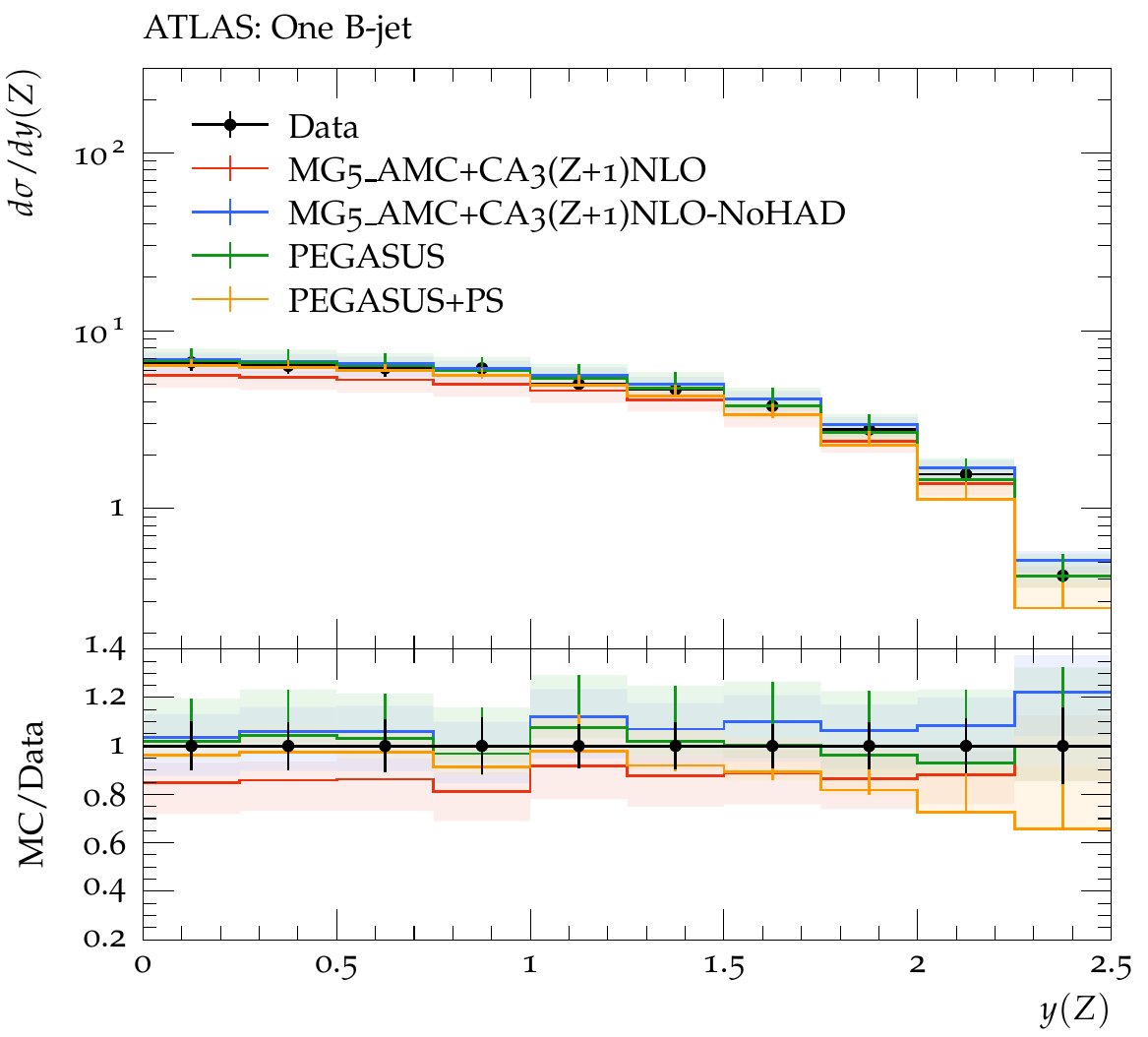}
\caption{$Z+b$ production differential cross sections as functions of $b$-jet 
and $Z$-boson transverse momenta and rapidities at $\sqrt{s}= 13$~TeV. ATLAS experimental data were taken from 
\cite{ATLAS13}.     
}
\label{fig2}
\end{center}
\end{figure}

\begin{figure}
\begin{center}
\includegraphics[width=7.8cm]{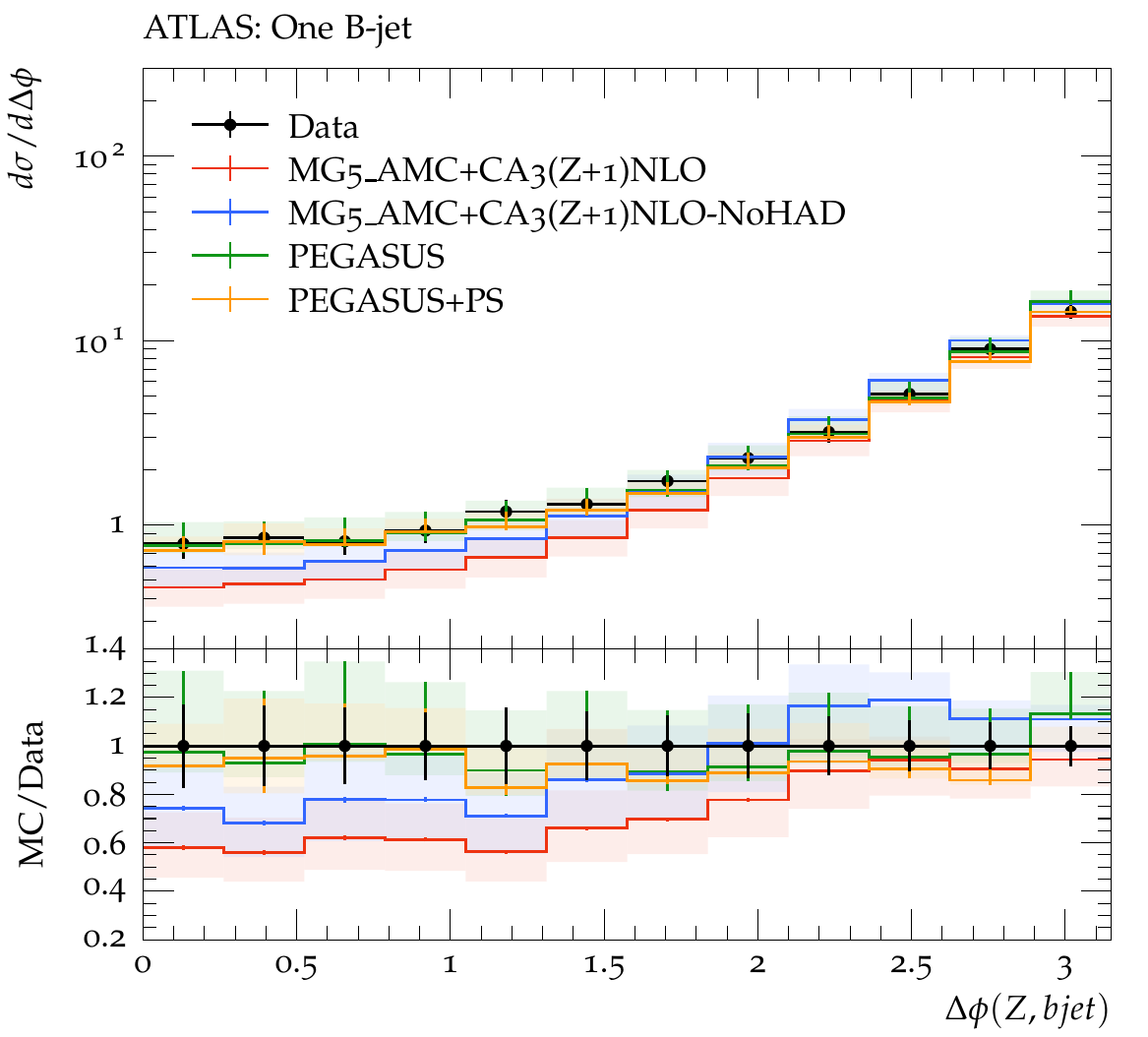}
\includegraphics[width=7.8cm]{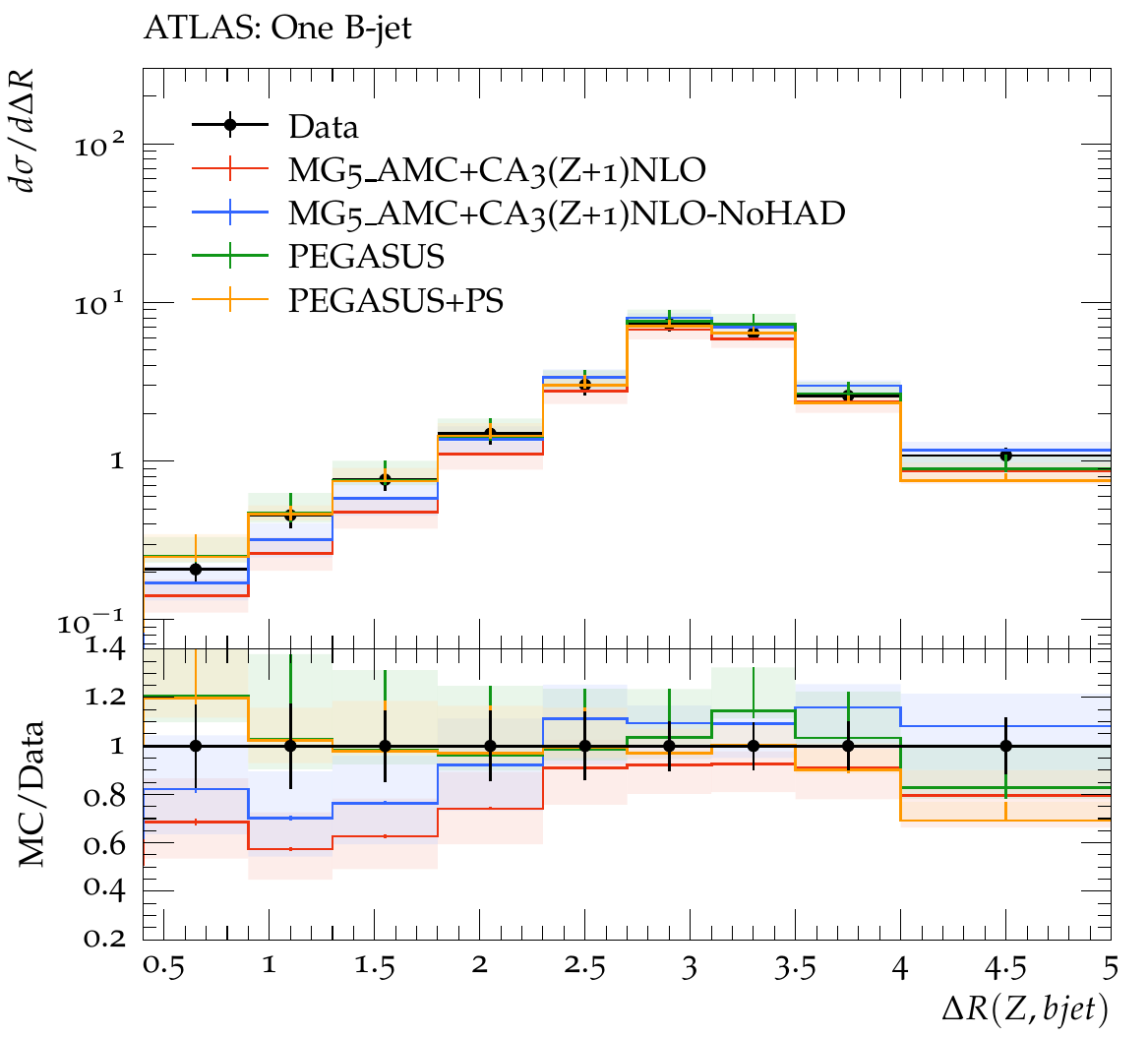}
\includegraphics[width=7.8cm]{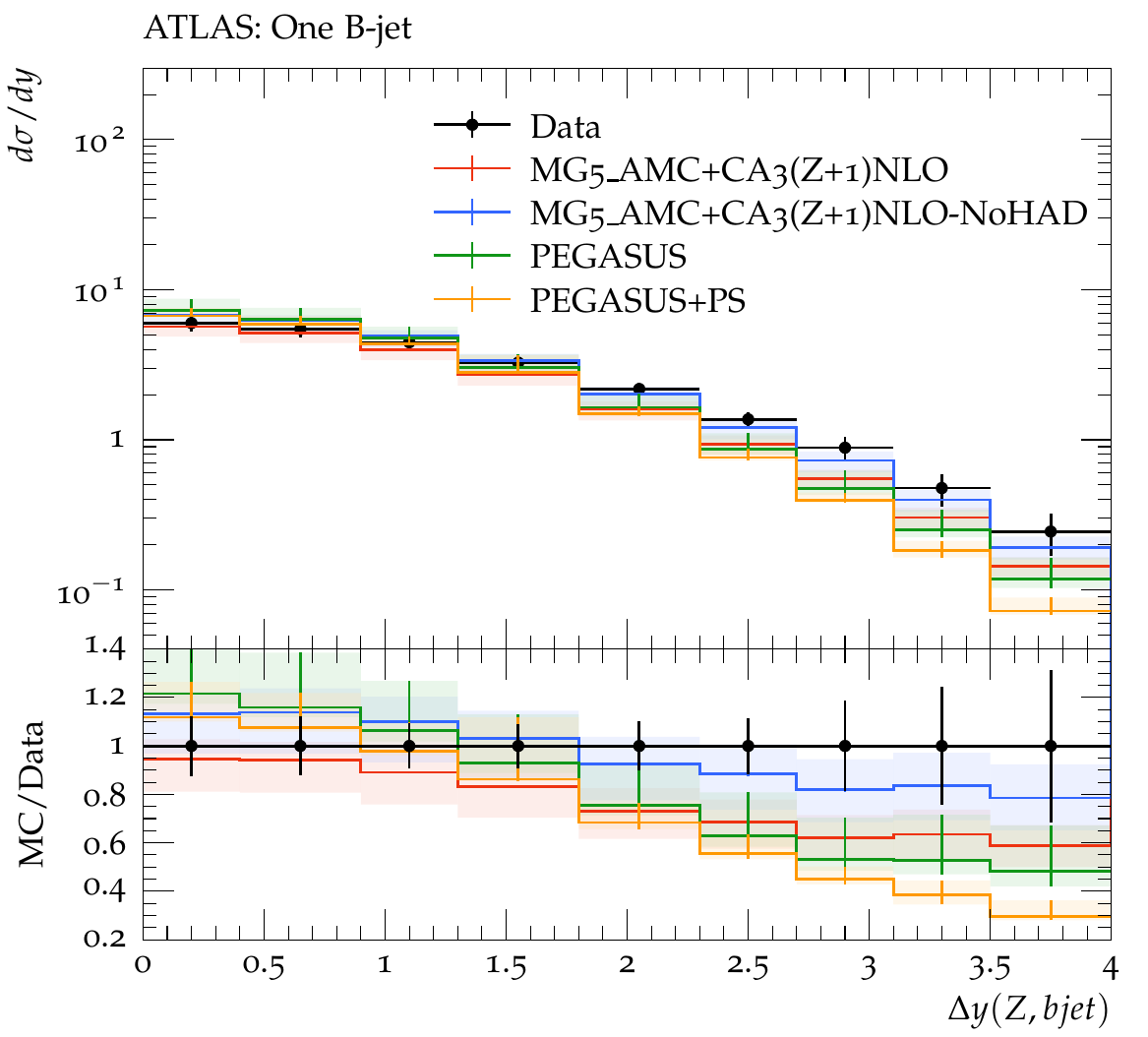}
\caption{$Z+b$ production differential cross sections as functions of the azimuthal angle, angular difference and rapidity differences between $b$-jet 
and $Z$-boson at $\sqrt{s}= 13$~TeV. ATLAS experimental data were taken from 
\cite{ATLAS13}.     
}
\label{fig3}
\end{center}
\end{figure}

\begin{figure}
\begin{center}
\includegraphics[width=7.8cm]{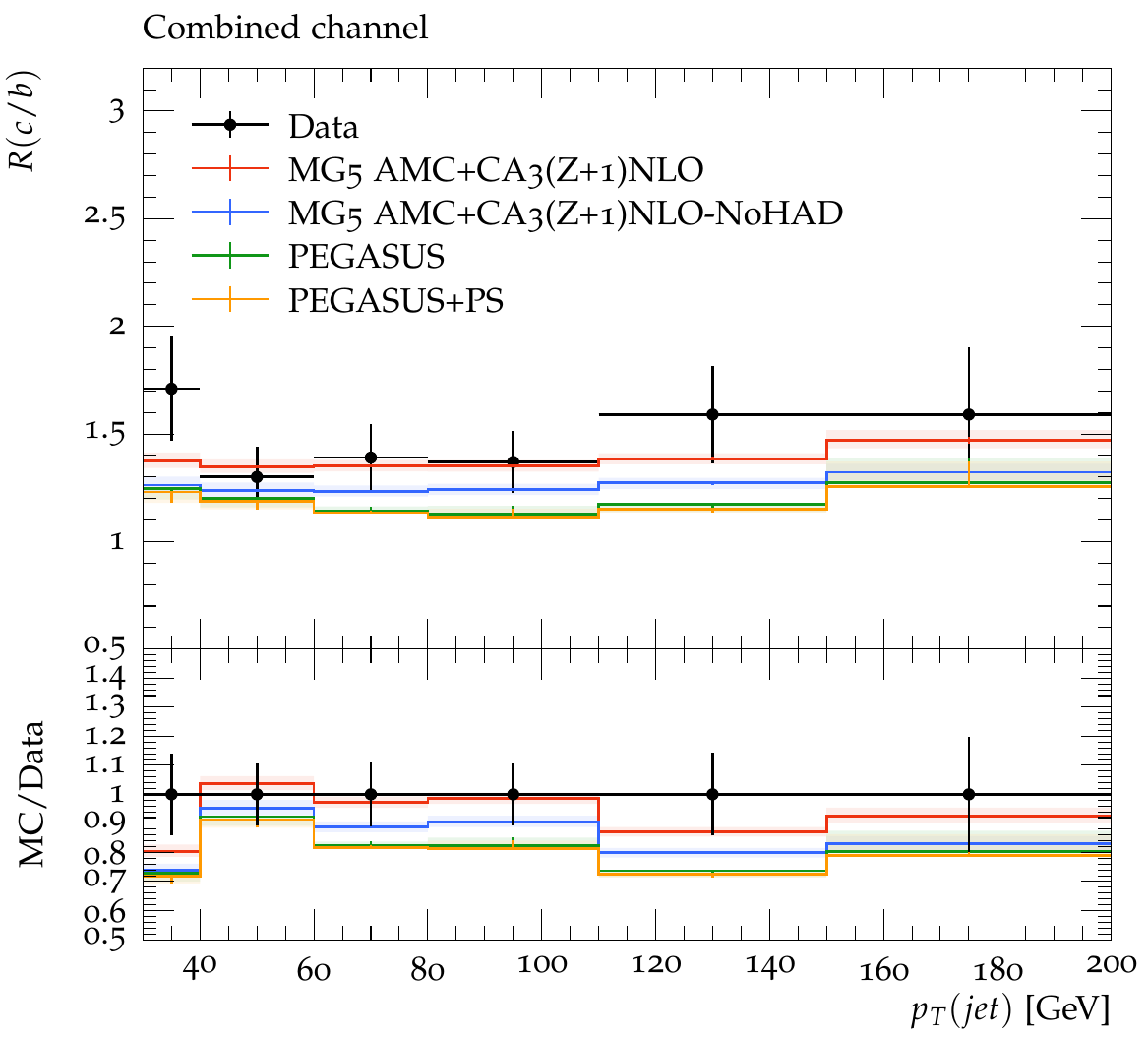}
\includegraphics[width=7.8cm]{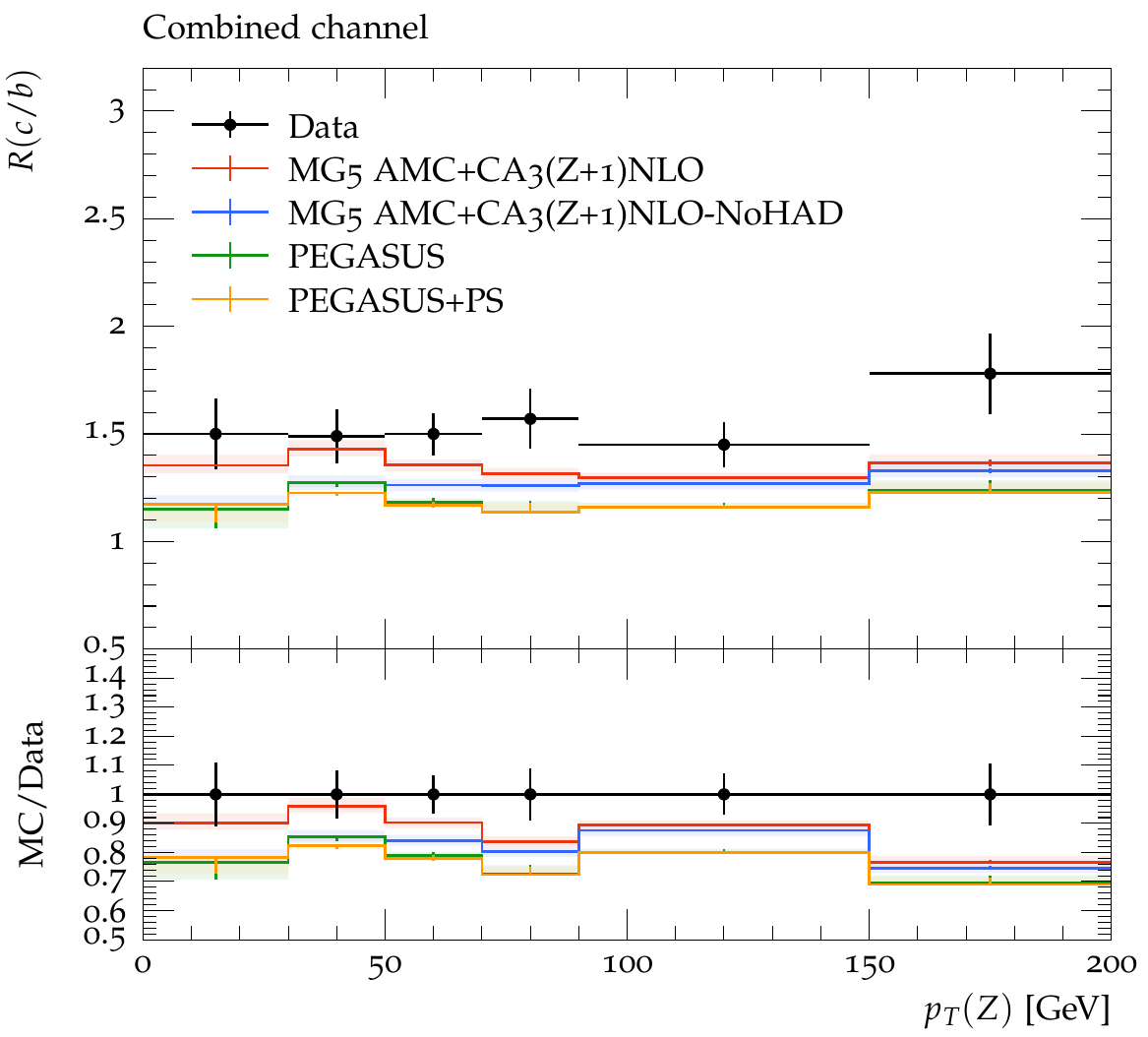}
\caption{The relative production rate $\sigma(Z + c)/\sigma(Z + b)$ as functions of heavy jet (left panel) 
and $Z$-boson (right panel) transverse momentum at $\sqrt{s}= 13$~TeV. 
The experimental data are from CMS\cite{CMS13}.}
\label{fig1}
\end{center}
\end{figure}

\subsection{Prompt and non-prompt $b$-jets} \indent

Since the two approaches considered above are suitable to describe latest LHC data on $Z + b$ production,
we turn to an investigation of observables sensitive to the different sources of heavy jets.
Here we concentrate mainly on $b$-jets, of course, the same arguments  can be applied for charmed jets. 

In the following we consider kinematic properties of $b$-jet production by investigating 
 $B$-hadrons  tagged via the semileptonic decay $B\to\mu+X$: 
we consider the fractional energy $z_b$ carried by the decay muon with respect to the total 
$b$-jet energy and the muon transverse momentum $p_T^{\rm rel}$ with respect 
to the $b$-jet axis. With both variables we aim to distinguish between prompt $b$-production, where
the $b$-quark exists already at the matrix element level, and non-prompt $b$-production, where the 
$b$-quark is generated during the jet evolution. We expect large $z_b$ and small $p_T^{\rm rel}$ (relative to the jet $p_T$)  for  
prompt $b$ production, while non-prompt $b$-production would lead to significantly smaller values for $z_b$ and a larger tail for $p_T^{\rm rel}$.

In Fig.~4  the distribution of  $z_b$ and $p_T^{\rm rel}$ are shown for different  thresholds of 
the jet transverse momentum: $p_T^{jets} >  30,\, 50,\, 100,\, 200,\, 300$~GeV.
Jets with larger transverse momenta  
provide larger phase space for parton radiation cascades. Accordingly, they show a larger
fraction of `non-prompt' $B$-hadrons resulting in a larger fraction of low-energy muons
(muons with low $z_b$) and in larger fraction of muons with high $p_T^{\rm rel}$
(muons with large deviation from the jet direction). 

We find, that the intuitive variables $z_b$ and $p_T^{\rm rel}$ are very powerful to distinguish prompt and non-prompt $b$ production

\begin{figure}
\begin{center}
\includegraphics[width=7.8cm]{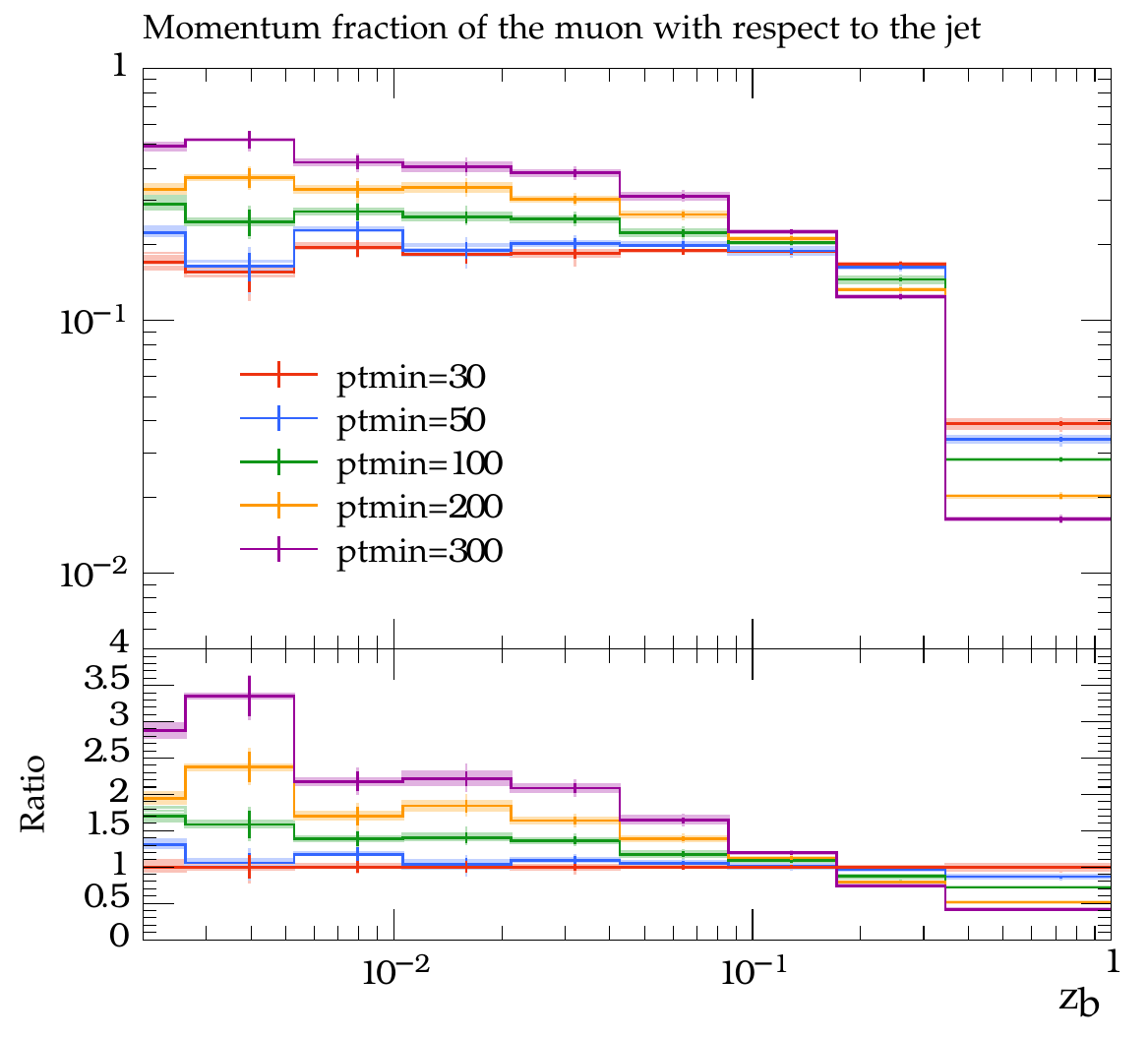}
\includegraphics[width=7.8cm]{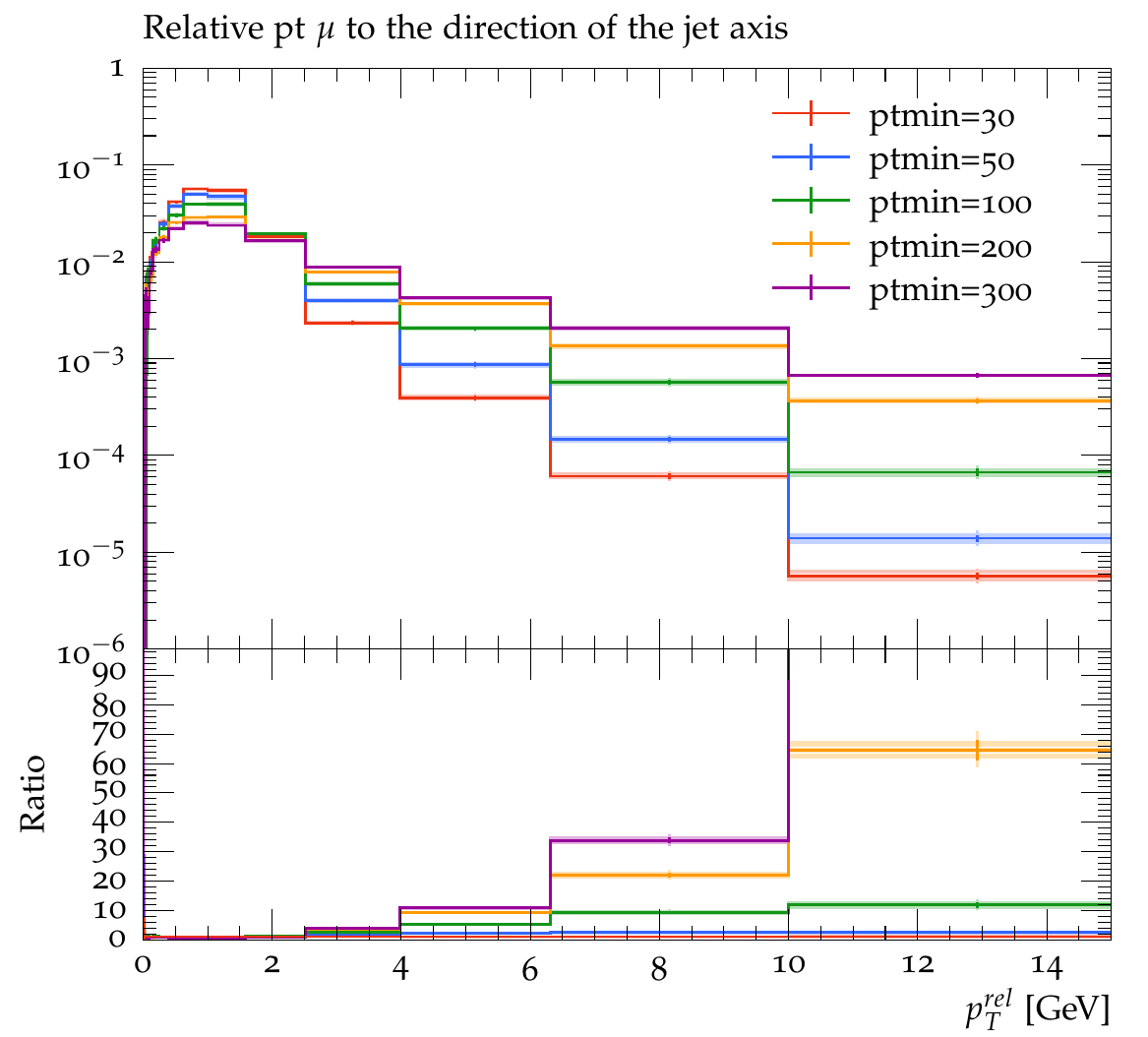}
\caption{$Z+b$ production differential cross sections as functions of the energy fraction carried by the decay muon with respect to the total 
$b$-jet energy and the muon transverse momentum with respect 
to the $b$-jet axis at $\sqrt{s}= 13$~TeV for different cuts on the minimal $p_T$ of the jet.     
}
\label{fig6}
\end{center}
\end{figure}

\section{Conclusion} \noindent

We have considered the production of $Z$ bosons associated with charm and beauty jets
at  LHC conditions. We investigated two different schemes.

We find that the combination of three basic subprocesses
(\ref{ggZ}) -- (\ref{qqsZ}) involving heavy quarks in final states 
provides a consistent description of $Z$ boson and/or $b$-jet transverse momenta and rapidity 
distributions as well as different correlation observables in $Z + b$ events at low and moderate $p_T$.
This can be seen from a direct comparison between the model predictions 
obtained using the Monte-Carlo generator \textsc{pegasus}
and recent LHC data.

In another approach we consider $Z+jet$ production at NLO, where heavy quarks can be produced
directly in the matrix element, or during the showering process. We perform the calculations based on
the Parton Branching TMDs together with TMD shower for the initial state cascade.
Such calculations were performed using \MGTMDZj . We find
very good description of the measurement at not too large transverse momentum of the $Z$ boson.

We classify different $b$-jet production mechanisms as prompt and non-prompt, depending whether
the heavy quark is present at matrix element level or generated during the jet evolution.

Events of the prompt and non-prompt types show rather different kinematic properties,
that can be seen, in particular, in relative production rate $\sigma(Z + c)/\sigma(Z + b)$
measured very recently by the CMS Collaboration for the first time.
The ratio of $c$ over $b$ jet production is also reasonably well described by the PB prediction of \MGTMDZj .

Considering the $Z + b$ events as a representative example,
we see that jets with larger transverse momenta contain larger fraction of non-prompt $b$-hadrons,
that results in larger fraction of low-energy muons and in larger fraction of muons with
large deviation from the jet direction. 
We recommend that the relevant observables, $z_b$ and $p_T^{\rm rel}$, be used in
the forthcoming experimental analyses.

\section*{Acknowledgments} \noindent

We thank G.I.~Lykasov and S.M.~Turchikhin for important comments and remarks.
S.P.B., A.V.L. and M.A.M. are grateful to
DESY Directorate for the support in the framework of
Cooperation Agreement between MSU and DESY on 
phenomenology of the LHC processes
and TMD parton densities.
M.A.M. was also supported by the grant of the Foundation for the Advancement of 
Theoretical Physics and Mathematics “BASIS” 20-1-3-11-1.
STM thanks the Humboldt Foundation for the Georg Forster research fellowship.

\end{document}